\documentclass[aps,prd,showpacs,preprintnumbers]{revtex4}
\usepackage{graphics}
\def \beq{\begin{equation}}
\def \eeq{\end{equation}}
\def \beqa{\begin{eqnarray}}
\def \eeqa{\end{eqnarray}}
\def \tr{{\rm tr}\,}
\def \re{{\rm Re}\,}

\def \det{{\rm Det}\,}
\def \x{{\mathbf x}}
\def \y{{\mathbf y}}
\newcommand{\cC}{{\cal C}}
\newcommand{\cD}{{\cal D}}
\newcommand{\cJ}{{\cal J}}
\newcommand{\cO}{{\cal O}}

\def \etal{{\sl et al.\/}}
\def \np{{\sl Nucl.\ Phys.\/}}
\def \pl{{\sl Phys.\ Lett.\/}}
\def \pr{{\sl Phys.\ Rev.\/}}
\def \prl{{\sl Phys.\ Rev.\ Lett.\/}}
\def \ie{{\sl i.\ e.\/}}

\begin{document}
 
\title{Chiral dynamics in QCD at finite chemical potential}
\author{Sourendu \surname{Gupta}}
\email{sgupta@theory.tifr.res.in}
\author{R.\ \surname{Ray}}
\email{rajarshi@theory.tifr.res.in}
\affiliation{Department of Theoretical Physics, Tata Institute of Fundamental
         Research,\\ Homi Bhabha Road, Mumbai 400005, India.}

\begin{abstract}

We present the renormalized Taylor expansion of the chiral condensate
with chemical potential. The relation with the Taylor coefficients of
meson susceptibilities and the quark mass dependence of the quark number
susceptibilities are obtained through chiral Ward identities and Maxwell
relations. The continuum limit is obtained in quenched and two flavour
QCD. In $N_f=2$ QCD, the quadratic response coefficients (QRCs) of the
chiral condensate to chemical potential indicate that pion fluctuations
decrease at finite chemical potential toward the critical end point.
We discuss how these observables can be used to test other features of
the QCD phase diagram.

\end{abstract}
\pacs{12.38.Aw, 11.15.Ha, 05.70.Fh\hfill
TIFR/TH/04-22, hep-lat/0409126\\ To appear in Physical Review D}
\maketitle

\section{Introduction}

Studies of hot dense matter formed in heavy-ion colliders such as the
RHIC \cite{rhic} need to be interpreted using theoretical predictions
obtained from lattice computations.  Most of these are performed at zero
baryon density. However, it seems likely that the actual experimental
conditions involve small but non-vanishing baryon chemical potential. In
order to deal with this one has to consider the stability of various
lattice predictions to the presence of a small chemical potential, $\mu$.

This was once considered an insoluble problem in lattice QCD. The
reason is that direct lattice Monte Carlo simulations at finite
$\mu$ are impossible since the weights in the partition function
are not positive definite. However, developments in recent years
\cite{fk,owe,pressure,biswa} mean that QCD at finite $\mu$ is no
longer out of reach on the lattice.  In this paper we investigate Taylor
expansions in the chemical potential \cite{gott,miya,pressure}, which
have recently been used to compute the equation of state at finite $\mu$
\cite{pressure,biswa}.

The first investigations of correlators at finite chemical potential was
performed in \cite{miya}. Our approach differs from theirs in that we
do a direct Taylor expansion of operators, rather than of parameters in
fitting functions.  There is a conceptual and practical gain.  Putting
chemical potentials on quarks can break the flavour$\times$CP symmetry
of QCD in many different ways, possibly affecting every observable in
distinct fashion. However it turns out that there are enough relations
between operators and their derivatives that, among all the observables
related to chiral dynamics in QCD, there is only one set of independent
Taylor coefficients.  Taylor expansions of the operators utilize these
relations with generality and economy.

There turn out to be three independent parameters of interest in the
chiral sector--- a single linear response coefficient of the chiral
condensate to $\mu$, and two quadratic response coefficients (QRCs).
The linear coefficient vanishes by symmetry in an expansion around
$\mu=0$. Thus, the two chiral QRCs encapsulate the physics arising from
the influence of baryon dynamics due to $\mu\ne0$ on chiral fluctuations.
Since the phase diagram of QCD at small $\mu$ is built up through the
interplay of chiral and baryon dynamics \cite{kr,ss,toublan,nishida},
the QRCs provide direct information on this phase diagram.

Here we present the first systematic study of the renormalized Taylor
expansion of the chiral condensate in QCD up to second order in the
chemical potentials \cite{miyappb}. We define renormalization schemes
which are simple to implement, and in which the Taylor expansion can
be extrapolated to the continuum. We find the continuum limits at
both high and low temperatures in quenched and dynamical $N_f=2$ QCD.

In section 2 we write out generic Taylor expansions in the chemical
potential, before specializing to pure glue operators, the chiral
condensate and meson susceptibilities. The renormalization schemes for the
condensate, Maxwell relations and chiral Ward identities are also dealt
with in this section. The actual lattice simulations and results are in
section 3, including the complete specification of the renormalization
schemes and results above and below $T_c$ in the quenched theory. This
section also contains our results for QCD with dynamical quarks.
Discussion of the results and the constraints that the QRCs can place
on the QCD phase diagram can be found in the concluding section 4.

\section{Taylor expansions}

The partition function of QCD at temperature $T$ and chemical
potentials $\mu_f$ for each flavour $f$ of quarks is
\beq
   Z(\{m_f\},T,\{\mu_f\})=\int\cD U {\rm e}^{-S(T)} \prod_f\det M(m_f,T,\mu_f),
\label{part}\eeq
where $m_f$ is the mass of quark flavour $f$. $S(T)$ denotes the gluon
part of the action and $M$ is the (lattice discretised) Dirac operator.
The expectation value of an operator $\cJ$ is defined as
\beq
   \langle\cJ\rangle\equiv J(\{m_f\},T,\{\mu_f\}) = \frac1Z
   \int\cD U \cJ {\rm e}^{-S(T)} \prod_f\det M(m_f,T,\mu_f),
\label{opex}\eeq
where $\cJ$ may or may not contain explicit dependence on the parameters.
Direct simulations are possible only along certain subspaces of the
full parameter space, such as $\re\mu_f=0$ or $\mu_u=-\mu_d$. However,
Taylor expansions around $\mu_f=0$ have proved to be a good method of
studying the physics for general $\mu_f$. In the remainder of this
paper we shall work mainly with the two flavour case $f\in\{u,d\}$
and with degenerate quark masses $m_u=m_d=m$. We shall freely go from
the parameter space expressed in terms of $\mu_u$ and $\mu_d$ to the
isoscalar chemical potential $\mu_0=(\mu_u+\mu_d)/2$ and the isovector
$\mu_3=(\mu_u-\mu_d)/2$. These are simply related to the baryon and
electric charge chemical potentials. Since we perform Taylor expansions
around the point $\mu_f=0$, all angular brackets in the rest of this
paper shall mean expectation values at vanishing chemical potential. These
are computable by the standard methods of lattice gauge theory.

We recall a few points about the expansion of the partition function
\cite{pressure,biswa}.  The $\mu_f$ appear only in the
Fermion determinant. If a matrix $M(x)$ depends on a parameter $x$, then
the identity $\det M(x)=\exp[\tr\ln M(x)]$ yields, $[\det M(x)]'=\det
M(x) \tr M^{-1} M'$, where primes denote derivative with respect to the
parameter $x$. Furthermore, the identity $M^{-1}M = 1$ yields $[M^{-1}]'=-
M^{-1}M'M^{-1}$.  This is enough to determine the Taylor coefficients of
the partition function.  The first order coefficients are proportional
to the quark number densities---
\beq
  \left.\frac{\partial Z}{\partial\mu_f}\right|_{\mu_f=0}\equiv ZZ_1,
         \qquad{\rm where}\qquad
     Z_1=\langle\cO_1\rangle=\langle\tr M_f^{-1}M_f'\rangle,
\label{ord1}\eeq
which is flavour independent since $m_u=m_d$. When we need to keep
track of the flavour we shall use the notation $\cO_1^{(f)}$ and obvious
generalizations for higher derivatives. The second order coefficients are
\beq
   \left.\frac{\partial^2Z}{\partial\mu_f\partial\mu_g}\right|_{\mu_f=0} = ZZ_2^{(fg)}
    =Z\langle\cO_{11}+\cO_2\delta_{fg}\rangle \qquad{\rm where}\qquad
   \langle\cO_2\rangle=-\langle\tr M_f^{-1}M_f'M_f^{-1}M_f'-\tr M_f^{-1}M_f"\rangle.
\label{ord2}\eeq
The notation used above is $\cO_{11}=\cO_1\cO_1$.
Flavour symmetry gives two independent components---
$Z_2^{uu}=Z_2^{dd}=\langle\cO_2+\cO_{11}\rangle$ and
$Z_2^{ud}=\langle\cO_{11}\rangle$. Expansions for isoscalar
($\mu_u=\mu_d=\mu_0$) and isovector ($\mu_u=-\mu_d=\mu_3$) chemical
potentials are---
\beq
Z(m,T,\mu_u,\mu_d)=Z(m,T,0,0)\times\left\{\begin{array}{l}
  1+\langle2\cO_{11}+\cO_2\rangle\mu_0^2+\cO\left(\mu_0^4\right),\\
  1+\langle\cO_2\rangle\mu_3^2+\cO\left(\mu_3^4\right).
  \end{array}\right.
\label{expand}\eeq
The terms of odd order vanish by CP symmetry. Diagrammatic representations of
these operators have been written down \cite{dgm} and can be extended to other
operators, as we outline later.

The double series expansion in terms of $\mu_u$ and $\mu_d$ of
the pressure, $P=T(\log Z)/V$, has as coefficients the quark number
densities, $n_f = (T/V) \langle \cO_1 \rangle = 0$, and the quark
number susceptibilities. The two independent linear susceptibilities
\cite{gott} are the diagonal ones $\chi_{uu} = \chi_{dd} = (T/V) \langle
\cO_{11}+\cO_2 \rangle$ arising from taking two derivatives with respect
to the same chemical potential, and the off-diagonal $\chi_{ud} = (T/V)
\langle \cO_{11} \rangle$. In writing these expressions we have used the
fact that the expansion is around $\mu_f=0$. Higher order susceptibilities
have also been used \cite{pressure}.

Similar Taylor expansions can be written for any operator $\cJ$. Here we write
this out to second order. Using the notation $\cJ^{(f)}$ for the first
derivative of $\cJ$ with respect to any explicit appearance of $\mu_f$ in it,
and $\cJ^{(fg)}$ for the explicit double derivative with respect to $\mu_f$ and $\mu_g$
we can write the first two Taylor coefficients in the expansion---
\beqa
\nonumber
   J_1^{(f)} &\equiv& \left.\frac{\partial J}{\partial\mu_f}\right|_{\mu_f=0} =
    \left\langle\cO_1^{(f)}\cJ+\cJ^{(f)}\right\rangle,\\
   J_2^{(fg)} &\equiv& \left.\frac{\partial^2J}{\partial\mu_f\partial\mu_g}\right|_{\mu_f=0} =
    \left\langle\biggl(\cO_{11}+\delta_{fg}\cO_2\biggr)\cJ
       +\cO_1\biggl(\cJ^{(f)}+\cJ^{(g)}\biggr)+\cJ^{(fg)}\right\rangle.
\label{dJ}\eeqa
The corresponding Taylor expansion is
\beqa
\nonumber
   J(m,T,\mu_u,\mu_d)&=&J_0+J_1^{(u)}\mu_u+J_1^{(d)}\mu_d+\\
      &&\qquad\frac12\left[\left(J_2^{(uu)}-J_0Z_2^{(uu)}\right)\mu_u^2
                  +\left(J_2^{(dd)}-J_0Z_2^{(dd)}\right)\mu_d^2
                  +2\left(J_2^{(ud)}-J_0Z_2^{(ud)}\right)\mu_u\mu_d\right]
       +\cdots
\label{deJ}\eeqa
where $J_0=J(m,T,0,0)$. CP symmetry cannot be used to simplify the
expansion in general, since the operator may not transform nicely
under CP. Special simplifications occur for purely gluon operators,
as we discuss later.

\subsection{Pure gauge operators}

Since pure gauge operators have no explicit dependence on the chemical
potentials,
\beq
   J_1 = \left\langle\cO_1\cJ\right\rangle,\quad{\rm and}\quad
   J_2^{(fg)} = \left\langle\biggl(\cO_{11}+\delta_{fg}\cO_2\biggr)\cJ\right\rangle.
\eeq
The corresponding Taylor expansion is
\beq
   J(m,T,\mu_u,\mu_d)=J_0+J_1(\mu_u+\mu_d)+
      \frac12\left[\left(J_2^{(uu)}-J_0Z_2^{(uu)}\right)(\mu_u^2+\mu_d^2)
                  +2\left(J_2^{(ud)}-J_0Z_2^{(ud)}\right)\mu_u\mu_d\right]+\cdots
\label{gauge}\eeq
In general the linear term vanishes exactly only when $\mu_0=0$. However,
for CP even operators, such as real parts of traces of products of links,
the odd order terms can be shown to vanish by summing over CP orbits
before performing the remaining part of the path integral.

\subsection{Chiral Condensate}

\begin{figure}[!tbh]
   \includegraphics{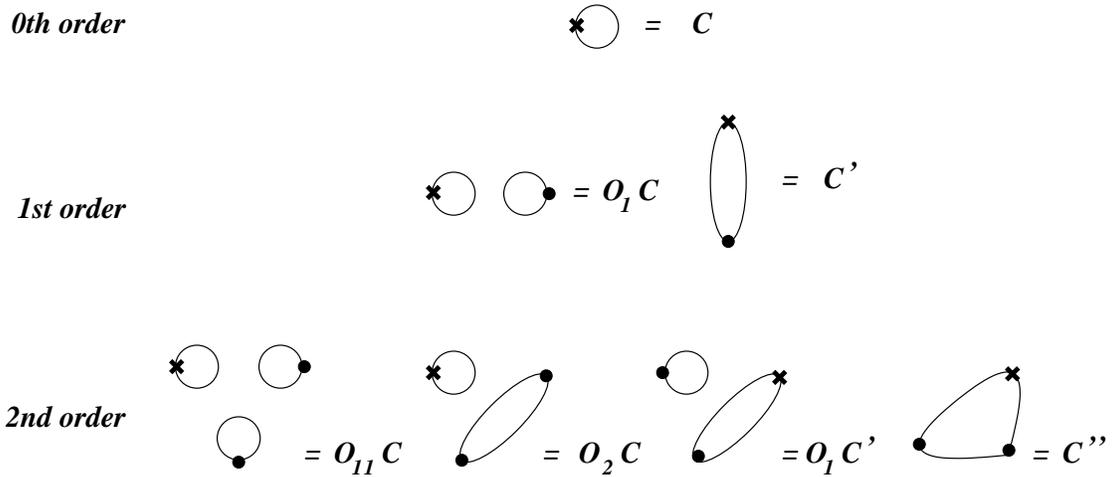}
   \caption{A diagrammatic representation of the terms in the expansion of the
      chiral condensate, extending
      the rules in \cite{dgm}. The lines represent quark propagators. The
      dots are insertions of $M'$ (which is $\gamma_0$ in the continuum
      limit) and the crosses are mass insertions. They represent derivatives
      with respect to $\mu$ and $m$ respectively. The contribution of each
      topology to the derivatives in eq.\ (\ref{ctaylor}) can be found by
      simple counting. Every closed loop is a Fermion trace. These diagrams
      should be thought of as being connected to all possible topologies of
      gluons.}
\label{fg.chicnd}\end{figure}

From the flavoured condensates $\langle\overline uu\rangle$ and 
$\langle\overline dd\rangle$, one can build the isoscalar and isovector
condensates which are, respectively, 
\beq
   C_S(m,T,\mu_u,\mu_d)=\frac12\left[
        \langle\overline uu\rangle+\langle\overline dd\rangle\right]
    \qquad{\rm and}\qquad
   C_V(m,T,\mu_u,\mu_d)=\frac12\left[
        \langle\overline uu\rangle-\langle\overline dd\rangle\right].
\label{condsv}\eeq
For equal mass quarks, $C_V$ vanishes at vanishing chemical
potential. This vector flavour symmetry is not broken by $\mu_0$ since
a sign flip in $\mu_0$ carries the quark of each flavour to its own
antiquark.  As a result, $C_V$ vanishes at all $\mu_0$, and the series
expansion for $C_S(\mu_0)$ is even.  A sign flip in $\mu_3$ carries
a quark of a given flavour to the antiquark of the opposite flavour,
thus mixing the vector flavour symmetry with CP. The same arguments can
still be followed through to show that $C_V$ vanishes to all orders in
the expansion around $\mu_3=0$ and $C_S(\mu_3)$ is even. In the remainder
of this paper we consider only $C_S$.

Recall that the quark condensate can also be written as
\beq
   C_S(m,T,\mu_u,\mu_f) = \frac1{2V_4}\left.
    \left(\frac{\partial\log Z}{\partial m_u}
    +\frac{\partial\log Z}{\partial m_d}\right)\right|_{m_u=m_d=m}
\label{cond}\eeq
where $V_4$ is the 4-volume of the lattice \cite{note}. Spontaneous breaking of
chiral symmetry is signaled by a non-zero value of the condensate
in the limit $m=0$. We shall not be concerned with this limit in
this paper, and can consequently neglect the known subtleties of taking
the limits $m\to0$ and $V_4\to\infty$. We shall however, check
that in actual numerical work the volumes are large enough to obtain
correct physical results.

Since the mass is subject to renormalization as the cutoff changes,
one must bother about cutoff dependence of the condensate as well
as its Taylor coefficients. Since we work with staggered quarks, the
mass renormalization is multiplicative, $m_R(a) = \sqrt Z_m(a) m$,
where the renormalization constant depends on the ultraviolet cutoff,
the lattice spacing $a$, but not on $T$ and $\mu$. Throughout this paper
we write the bare quark mass as $m$. The quark condensate
is also renormalized by $Z_m$.  If we are not interested in the absolute
value of the condensate, but only on its variation with $\mu$ at constant
$T$, then one can simply work with either of the ratios
\beq
   \frac{C_S(m_R,T,\mu_u,\mu_d;a)}{C_S(m_R,0,0,0;a)} \quad({\rm Z\ scheme})
      \qquad{\rm or}\qquad
   \frac{C_S(m_R,T,\mu_u,\mu_d;a)}{C_S(m_R,T,0,0;a)} \quad({\rm T\ scheme}).
\label{scheme}\eeq
One expects the T scheme to be undefined for $T>T_c$ when $m_R=0$,
since the chiral condensate then vanishes. Elsewhere there is a finite
multiplicative relation between the schemes which has the interpretation
of the $T$-dependence of $C_S$ at $\mu=0$. The computations have to be
performed at fixed renormalized mass. We perform a formal expansion here
which will later be renormalized in either of these schemes.

For the $u$ quark condensate $ \langle\cC_u \rangle = 
\langle \overline uu \rangle = \langle \re \tr M_u^{-1}
\rangle$, clearly $\cC_u^{(d)} = \cC_u^{(ud)} = \cC_u^{(dd)}
= 0$. The non-vanishing derivatives are $\cC_u^{(u)} =
-\re \tr M_u^{-1}M_u'M_u^{-1} = \cC'$, and $\cC_u^{(uu)} = \re
\tr(2M_u^{-1}M_u'M_u^{-1}M_u'M_u^{-1}-M_u^{-1}M_u''M_u^{-1}) = \cC''$.
For the $d$ quark condensate, the corresponding relations can be found
by flipping $u$ and $d$ indices. Since the Taylor series expansion is
carried out at $\mu=0$, in the limit of degenerate masses the quantities
$\cC'$ and $\cC''$ are flavour independent.  In this limit the Taylor expansion
coefficients of the flavoured condensates are flavour independent, and we
need to evaluate the four quantities---
\beqa
\nonumber
   C_S^0 &=& \left\langle\cC\right\rangle,\\
\nonumber
   C_S^1 &=& \left\langle2\cO_1\cC+\cC'\right\rangle,\\
\nonumber
   C_S^{20} &=& \langle2(\cO_{11}+\cO_2)\cC+2\cO_1\cC'+\cC''\rangle
             -2\langle\cO_{11}+\cO_2\rangle\,\langle\cC\rangle,\\
   C_S^{11} &=& 2\langle\cO_{11}\cC+\cO_1\cC'\rangle
             -2\langle\cO_{11}\rangle\,\langle\cC\rangle,
\label{ctaylor}\eeqa
A direct check shows that $C_S^1$ vanishes, as does every odd term in the
expansion.  $C_S^{20}$ is the diagonal, and $C_S^{11}$
is the off-diagonal QRC.

A diagrammatic representation of the expansion of the chiral condensate
is shown in Fig.\ \ref{fg.chicnd}. This builds on a previous diagrammatic
representation of the expansion of the partition function \cite{dgm}. Every line
denotes a fermion propagator, a dot denotes a derivative with respect
to $\mu_f$ (an insertion of $\gamma_0$ in the continuum), every cross is
a derivative with respect to $m_f$ (a mass insertion), and every closed
line is a trace over Fermion coordinates. The fermions have to be dressed
in all possible ways by gluon insertions. Power counting arguments
\cite{bir} can be easily extended to estimate the power of the gauge
coupling $g$ associated with the connected parts which appear in eqs.\
(\ref{ctaylor}) when they are evaluated at high temperatures.  The largest
quark-loop-disconnected quantity is $\langle\cO_2\cC\rangle_c$ which is of
order $g^4$. Both $\langle\cO_{11}\rangle$ and $\langle\cO_1\cC'\rangle_c$
are of order $g^6$, and $\langle\cO_{11}\cC\rangle_c$ is of order
$g^{8}$. These powers may be modified by $\log g$ terms due to infrared
singularities. $\langle\cC''\rangle$ is of order $g^0$, and, because it
is non-vanishing, generically gives the largest contribution in the high
temperature phase.

The Taylor expansions of $C_S$ are
\beqa
\nonumber
   C_S(m,T,\mu_0) &=& C_S^0+\left(C_S^{20}+C_S^{11}\right)\frac{\mu_0^2}2+\cdots\\
   C_S(m,T,\mu_3) &=& C_S^0+\left(C_S^{20}-C_S^{11}\right)\frac{\mu_3^2}2+\cdots
\label{conds}\eeqa
where the temperature dependence of the coefficients has not been
shown explicitly.  Note that in the T scheme the Taylor series begin
with a constant term of exactly unity.  In this scheme one can rewrite
the series as an expansion in $\mu/T$ and render all the coefficients
dimensionless.  Then the remaining Taylor coefficients have the
significance of fractional changes due to non-zero values of $\mu$.
The power counting argument shows that for $T\gg T_c$ the coefficient of
the quadratic term is dominated by the operators $\langle\cC''\rangle_c$,
and the two expansions differ only by terms subleading in $g$. In the
low-temperature phase these arguments fail, and the two series could have
significantly different behaviour. Near and below $T_c$ other arguments,
outlined in Section \ref{results}, also lead us to expect different
behaviour for the two series.

\subsection{Chiral Ward identities for meson susceptibilities
\label{corr}}

A meson operator is $M(\x)=\overline\psi(\x)\Gamma\psi(\x)$, where
$\psi$ is a quark field operator and $\Gamma$ is a Hermitean
spin-flavour matrix. The conjugate operator is $M^\dag(\x) =
\psi^\dag\Gamma^\dag\gamma_0\psi = \overline\psi\widetilde\Gamma\psi
=\pm\overline\psi\Gamma\psi$, where the plus sign is obtained for the
Dirac matrices $1$, $\gamma_0$ and $\gamma_5\gamma_0$ and the negative
sign for $\gamma_5$, $\gamma_i$ and $\gamma_5\gamma_i$.  The meson
correlators are
\beq
   C(\x,\y) = \langle M(\x)M^\dag(\y)\rangle = 
    \langle\tr\Gamma M^{-1}\Pi(\y)\widetilde\Gamma(M^\dag)^{-1}\Pi(\x)\rangle,
\label{prop}\eeq
where $M$ is the Dirac operator, and $\Pi(\x)$ is the projector to
one site $\x$. The final expression on the right was obtained using
the relation $\Pi^2=\Pi$ and the fact that $\Pi$ and $\Gamma$ commute
since they act on different spaces.  In \cite{miya} the correlator was
parametrized in the form $A\cosh m(L/2-x)$ and the derivatives with
respect to $\mu$ of the parameters $A$ and $m$ were related to operator
expectation values obtained by taking derivatives of the right hand
side of eq.\ (\ref{prop}). However, since CP symmetry is explicitly
broken on introduction of chemical potentials, one need no longer have
$C(\x,\y) = C(|\x-\y|)$. This is explicitly visible in the computations
reported in \cite{myisov,hands}.

A little tinkering does not help, since it is the transfer matrix formalism
that needs to be re-examined at finite $\mu$, because a non-zero value of this
parameter introduces an explicit breaking of time reversal symmetry. It is
much more straightforward to deal with the meson susceptibilities \cite{old},
\beq
   \chi_\Gamma = \frac1V\sum_{\x\y}C(\x,\y)
    =\frac1V\langle\tr\Gamma M^{-1}\widetilde\Gamma(M^\dag)^{-1}\rangle.
\label{susc}\eeq
If a single pole dominates this propagator then the corresponding mass,
$m_\Gamma\propto1/\sqrt{\chi_\Gamma}$. However, this quantity more
generally measures mesonic fluctuations \cite{old}.

Chiral Ward identities are consequences of certain operator equalities
which follow from chiral symmetries. Each formulation of lattice
fermions can preserve a different subset of these identities; a basic
set for staggered fermions can be found in \cite{toolkit}. The remaining
continuum identities are violated at finite lattice spacing, and hence can
be used to check the approach to the continuum. The prototypical chiral
Ward identity is $C_S(T,\mu) = m \chi_\pi(T,\mu)$, where $\chi_\pi$
is the pseudo-scalar susceptibility.  Furthermore, this
chiral Ward identity allows us to define a renormalized $\chi_\pi$ using
the simultaneous renormalization of the quark mass and the condensate.
Taylor expansions of both sides of this identity can then be equated
term by term.  Another conclusion is that the Taylor expansion of
$\chi_\pi$ must be even in $\mu$. Explicit construction of the first
Taylor coefficient shows that it vanishes, in conformity with this proof.

A second chiral Ward identity \cite{patel} relates the isovector scalar
susceptibility to the mass derivative of the condensate: $\partial C_S/
\partial m= -\chi_\epsilon$. Again this can be used to renormalize
$\chi_\epsilon$ and relate two Taylor series.  We emphasize that this
involves the isovector scalar, \ie, the scalar meson susceptibility
implicated here is obtained out of a connected quark loop.

We show later that in the continuum limit of the high temperature phase
the second derivatives of the condensate with respect to the chemical
potential vanish.  As a result, the QNS are insensitive to variations
of the quark mass, and the pion susceptibilities are insensitive
to chemical potential. Also, in this phase, symmetry arguments show
that $\chi_\pi=\chi_\epsilon$ \cite{symmetry}, implying that $C_S$
vanishes linearly with $m$. As a result, we can deduce that the scalar
susceptibility is also independent of the isoscalar chemical potential, at
least to quadratic order. Below $T_c$ this chain of logic does not hold.
The scalar susceptibility is interesting because of speculation about the
massless modes at the critical end point \cite{hatta}. We will discuss
it at greater length elsewhere.

\subsection{Maxwell relations}\label{sc.maxwell}

A Maxwell relation is the equality of two distinct physical
interpretations of a mixed derivative obtained by interchanging the order
of the derivatives. From the Taylor expansion of the chiral condensate
in eq.\ (\ref{conds}) we can find Maxwell relations with the change of
quark number susceptibilities with the quark mass. The leading order
relation
\beq
    \frac{\partial\langle\overline\psi\psi\rangle_S}{\partial\mu}
       = \frac{\partial n}{\partial m}
\label{leading}\eeq
was first noted in \cite{kogut}. It is trivially true at $\mu=0$ since
the first derivative on the left vanishes, as does $n$ for all quark
masses.

The second derivatives give two non-trivial Maxwell relations and
consequent relations using the chiral Ward identities discussed earlier,
\beqa
\nonumber
    C_S^{20} &=& 
    \frac{\partial^2\langle\overline\psi\psi\rangle_S}{\partial\mu_u^2}
     = m\frac{\partial^2\chi_\pi}{\partial\mu_u^2}
       = \frac{\partial\chi_{uu}}{\partial m},\\
    C_S^{11} &=& 
    \frac{\partial^2\langle\overline\psi\psi\rangle_S}{\partial\mu_u\partial\mu_d}
     = m\frac{\partial^2\chi_\pi}{\partial\mu_u\partial\mu_d}
       = \frac{\partial\chi_{ud}}{\partial m}.
\label{maxwell}\eeqa
There are higher order relations between the mass derivative of the
non-linear susceptibilities and the higher Taylor coefficients of the
condensate.  This class of Maxwell relations can be checked by explicit
differentiation.  Preliminary results have been reported in \cite{nara}
and a mixed version of this has been used later to correct for the mass
dependence of a Taylor expansion of the pressure \cite{maria}. Here,
as a byproduct of our computation of the renormalized chiral condensate,
we shall give the continuum limit of the derivative of the susceptibility.

The relative rates of strange and light quark production in heavy-ion
collisions \cite{conti} is
\beq
   \lambda_s = {\chi_{ss} \over \chi_{uu}},
\eeq
with obvious extensions to the production rates of heavier quarks. Since
it is hard to perform lattice computations at realistic values of light
quark masses due to constraints of computer time, one can lighten the
computational burden by using a Taylor series for the mass dependence of
the susceptibilities utilizing the above Maxwell relation. Then one can
compute $\lambda_s$ at some reasonably light quark mass, corresponding
to, say, the pion mass being two to three times heavier than in the
real world, and extrapolate to the physical quark mass values using the
Maxwell relation. One can also investigate the stability against small
variations in the heavy quark mass by similar means.

\section{Results}

\subsection{Implementing the renormalization scheme}

\begin{table}[!htbp]
  \begin{center}\begin{tabular}{|c|c|c|c|c|c|c|c|c|c|c|}  \hline
  $a T_c$ &$\beta$ & size   & $am_q$  & $a^2m_{\pi}^2$ & $\chi^2$ & Range & $am_
{\rho}$ & $\chi^2 $ & Range & $m_{\pi}/m_{\rho}$ \\
  \hline
  0.1667  &5.8941  & $16^3 \times 32$ & 0.01667 & 0.119(1) & 1.99 &  7-10  & 0.66(2) &1.43&5-17&0.52(2) \\
          &        &                  & 0.02    & 0.141(2) & 2.82 &  7-10  & 0.68(2) &1.72&5-17&0.55(2) \\
          &        &                  & 0.03    & 0.205(2) & 1.39 &  8-10  & 0.74(2) &2.54&5-17&0.61(2) \\
          &        & $30^4$           & 0.01667 & 0.123(3) & 4.41 &  6-11  & 0.71(4) &10.73&3-16&0.49(3) \\
          &        &                  & 0.02    & 0.142(2) & 5.59 &  6-10  & 0.72(3) &11.08&3-16&0.52(2) \\
          &        &                  & 0.03    & 0.214(4) & 4.10 &  6-11  & 0.77(3) &11.78&3-16&0.60(2) \\
  \hline
  0.1333  &6.0150  & $20^3 \times 40$ & 0.01333 & 0.080(1) & 1.66 &  7-13  & 0.56(2) &4.95&5-21&0.50(2) \\
  \hline
  0.1111  &6.14    & $24^3 \times 48$ & 0.01111 & 0.0515(9)& 1.64 &  9-13  & 0.42(1) &2.62&7-25& 0.54(1) \\
  \hline
  0.0833  & 6.3384 & $40^4$           & 0.0083  & 0.035(1) & 6.58 &  9-17  & 0.333(9)&2.55&7-21& 0.56(2) \\
          &        &                  & 0.01    & 0.040(1) & 5.06 & 10-17  & 0.341(8)&2.64&7-21& 0.59(2) \\
  \hline
  0.0667  & 6.525  & $42^4$           & 0.0067  & 0.022(1) & 2.45 & 13-19  & 0.25(1) &7.35&8-22& 0.60(3)   \\
          &        &                  & 0.005   & 0.018(1) & 2.85 & 13-19  & 0.25(1) &6.67&8-22& 0.54(3)   \\
  \hline
  0.0556  &6.65    & $50^4$           & 0.0056  & 0.0161(3)& 8.13 & 13-23  & 0.21(1) &15.0&10-25& 0.60(3)  \\
  \hline
  \end{tabular}\end{center}
  \caption[dummy]{Meson masses in quenched QCD with Wilson action and staggered
     quarks. The $\chi^2$ for pions is obtained by fitting local masses to a
     constant. The rho meson mass is obtained with a four parameter fit to the
     vector correlator. Details are discussed in the text.}
\label{tb.runs}\end{table}

\begin{figure}[!tbh]
   \includegraphics{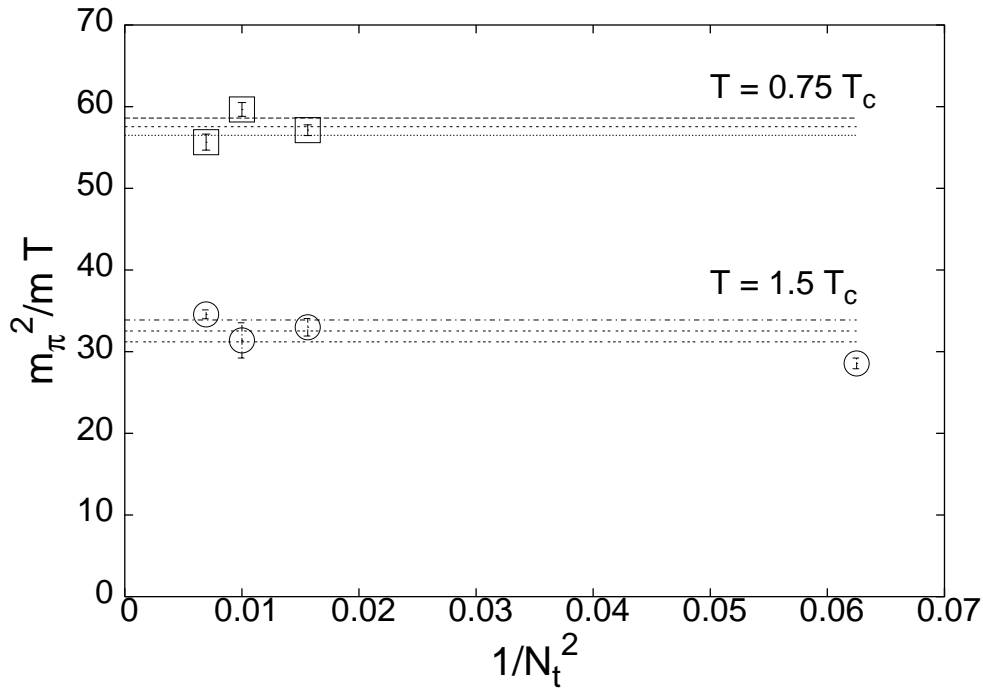}
   \caption{The dimensionless ratio $m_\pi^2(T=0)/mT$ for fixed bare quark
      mass $m/T_c=0.1$ at a succession of lattice spacings corresponding to
      fixed $T=1.5T_c$ at varying $N_t$ shown as a function of $1/N_t^2 =
      a^2T^2$. The constancy of the ratio shows that this procedure correctly
      holds the renormalized quark mass constant and thereby implements the
      renormalization scheme of eq.\ (\protect\ref{scheme}).}
\label{fg.renorm}\end{figure}

\begin{figure}[!tbh]
   \includegraphics{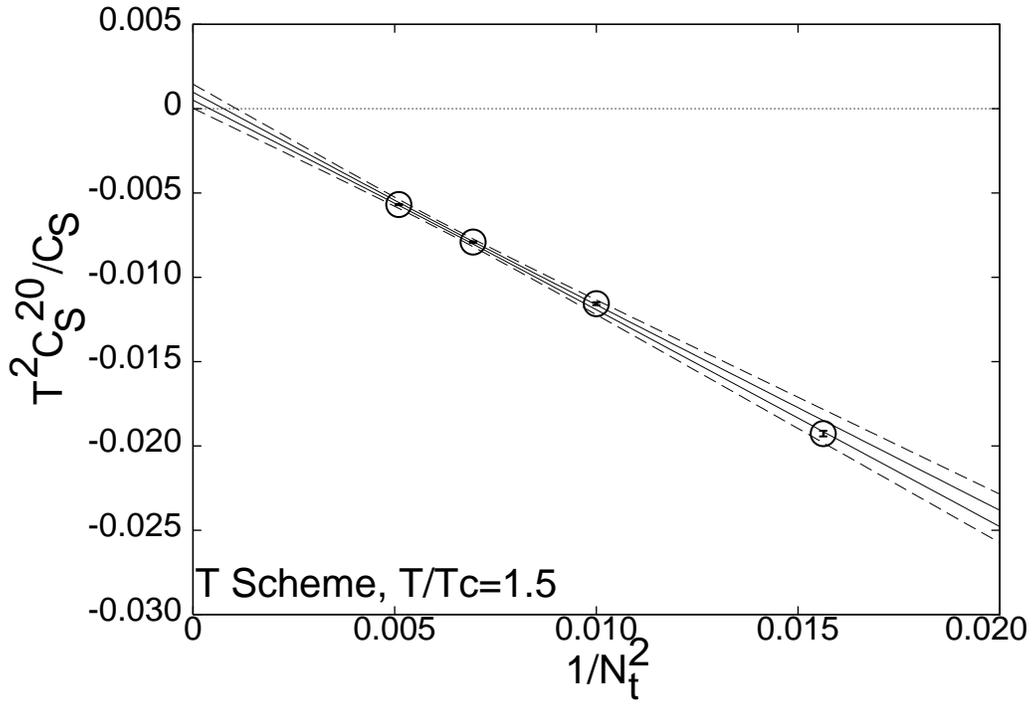}
   \caption{The diagonal QRC in the T scheme, at fixed $T/T_c=1.5$
            holding the renormalized quark mass fixed by fixing the
            bare mass to be $m/T_c=0.1$. The full lines show the
            1-$\sigma$ band on the continuum extrapolation and the
            dashed lines show the 3-$\sigma$ confidence band.}
\label{fg.scnord}\end{figure}

\begin{figure}[!tbh]
   \includegraphics{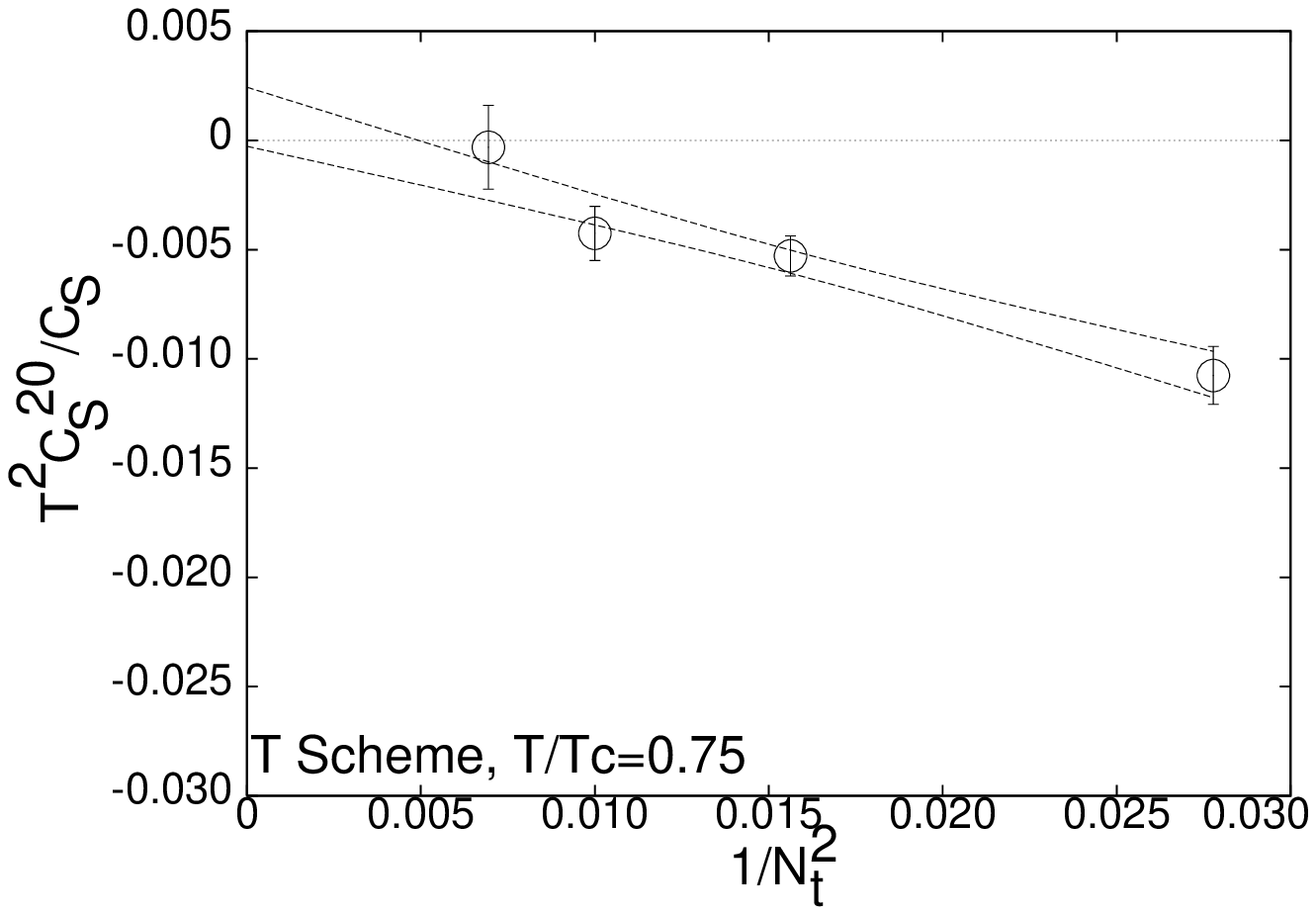}
   \caption{The diagonal QRC in the T scheme, at fixed $T/T_c=0.75$
            holding the renormalized quark mass fixed by fixing the
            bare mass to be $m/T_c=0.1$. The lines show the 1-$\sigma$
            band on the continuum extrapolation.}
\label{fg.scnorc}\end{figure}

In eq.\ (\ref{scheme}) we have defined a renormalization scheme in
which the multiplicative renormalization of the chiral condensate at
temperature $T$ and chemical potential $\mu$ is absorbed by the bare
condensate at $T=0$ or $\mu=0$ evaluated at the same lattice spacing $a$.
In order to implement this scheme this compensation must be performed
at a constant renormalized quark mass. We complete the renormalization
prescription by choosing to trade the pion mass for the renormalized
quark mass. The practical question is then the choice of the bare quark
mass which corresponds to a fixed pion mass.

The lattice spacing has been connected to the temperature scale using the
QCD $\beta$-function at full two-loop order \cite{scale}.  This scale
determination was used to tune the bare couplings in the Wilson action
required to simulate the quenched theory at a fixed physical temperature
while changing the lattice spacing \cite{valence}. Here we used the same
bare couplings to extract the pion mass using $T=0$ simulations.

The series of bare couplings used correspond to fixed $T=1.5T_c$ on
lattices with $N_t=4$, 8, 10 and 12 and to fixed $T=0.75T_c$ for
$N_t=8$, 10 and 12. The lattice sizes, $L$, were
selected so that $Lm_\pi\ge6$ in order to keep finite volume effects under
control.  At the two coarsest lattice spacings a previous determination
of meson masses \cite{rajan} could be used to estimate $m_\pi$ and
select an appropriate value of $L$. Then one could successively use the
already determined masses to estimate the size which would be required
at finer lattice spacing. Quenched configurations were generated using
the Wilson action.  At each lattice spacing the first 500--1000 sweeps
were discarded for thermalisation and a configuration was kept after
every subsequent 50 sweeps.

The stored configurations were used to determine the chiral condensate,
through a noisy evaluation of the trace of the inverse of the fermion
matrix, and meson correlators. The stopping criterion on the staggered
fermion matrix inversion was decreased by factors of 10 until the
extracted value of $m_\pi$ was seen to stabilize. It was observed that
while decreasing lattice spacing keeping $m_\pi$ fixed the stopping
criterion had to be decreased significantly. We attribute this to the fact
that as the lattice spacing decreases at zero temperature the 
eigenvalues near a rapidly decreasing renormalized quark mass take
more conjugate gradient steps to resolve.

Our extraction of pion masses was based on effective (local) mass
extraction, $m(t)$, and cross checked by fits.  As has been known for a
long time, this is contaminated at small distances by higher lying states
and at large distances by periodic lattice artifacts.  When a plateau can
be seen, that value is taken as an estimate of the ground state mass. We
converted this into an automatic task by varying $t_{\rm min}$ and $t_{\rm
max}$ and for each fixed value of this parameter checking whether the data
on $m(t)$ could be fitted to a constant. We accepted the largest interval
over which this could be done and took the corresponding constant as an
estimate of the mass (see Table \ref{tb.runs}. For the vector meson the
local mass is harder to stabilize, and we used four parameter fits to
extract the mass, varying the fit interval to look for stability.

In simulations at the coarser lattice spacings we performed a series
of computations at larger bare quark mass. These are useful in order
to check that the lattice volumes are under control. The near constancy
of the ratio $a^2m_\pi^2/(a m_q)$ at fixed $a$ on the coarser lattices
indicates that the lattice volumes are sufficiently large that one can
extrapolate to the correct continuum physics.

In simulations at a sequence of lattice spacings, $a$, corresponding to
a fixed temperature and varying $N_t$ ($a=1/TN_t$), with a sequence of
bare quark masses $m/T_c=0.1$, we found that the dimensionless ratio
$m_\pi^2/mT$ is fixed. This is shown in Figure \ref{fg.renorm}. The
constancy of this ratio shows that this sequence of bare quark masses
corresponds to a fixed renormalized quark mass, and gives an easy method
to tune the bare quark mass in order to fix the renormalized quark
mass. At this quark mass, we found that $m_\pi/m_\rho\approx0.6$,
indicating that $m_\rho$ could also have been chosen to fix the
renormalized quark mass. For the parameters chosen here, the fact that
the renormalized quark mass is rather heavy is seen both in the fact
that the ratio $m_\pi/m_\rho$ is larger than the physical value and
also in that $m_\pi/T_c\approx2.3$. We recall that with two flavours
of dynamical staggered quarks choosing $m/T_c=0.1$ corresponds to
$m_\pi/m_\rho=0.36\pm0.01$, and that on reducing the bare quark mass
to $m/T_c=0.03$ the ratio $m_\pi/T_c$ became physical \cite{su3}.

\subsection{Continuum limit in quenched QCD}
\subsubsection{$T>T_c$}

We used stored gauge configurations from the study in \cite{valence}
for our measurements. These were obtained on $N_t =
4, 8, 10, 12 $ and $14$ lattices for temperatures $T = 1.5 T_c, 2 T_c $
and $3 T_c$ respectively. The main concern in choosing the spatial size
$N_s$ is to avoid gross distortions of the infinite volume result. For
this it is sufficient to have $N_s/N_t>max(2,T/T_c)$. For $N_t = 4$
we have used $N_s = 12$. For all the other $N_t$ the spatial size was
chosen to be $N_s = 2 N_t + 2$ for $T/T_c=1.5$ and 2, and $N_s = 3 N_t + 2$
for $T/T_c=3$. For details of the generation of the configurations and
their statistical properties, we refer the reader to \cite{valence}.
In this work we recomputed, as a check, the operators $\cO_1$, $\cO_{11}$
and $\cO_2$ which were studied in \cite{valence} and supplemented
these by computations of $\cC$, $\cC'$ and $\cC''$ needed for the
computation of the Taylor expansion of the chiral condensate. The traces
were evaluated by the noisy method (see \cite{valence} for details).
For the conjugate gradient algorithm to obtain the matrix inverses,
we used a stopping criterion that the norm of the residual should be
less than $10^{-3}\sqrt{N_tN_s^3}$.

The vanishing of the expectation values $\langle\cO_1\rangle$ and
$\langle\cC'\rangle$ are checks of the accuracy of the computation. On
each lattice and at every temperature, we found that the expectation
values were zero to within a part in $10^6$ with the above stopping
criterion. In the high temperature phase of QCD, where we work,
$\chi_{ud}$ is small, and therefore one expects a noisy and almost
vanishing signal for its derivative with respect to the quark mass. In
agreement with this expectation, we find that $C_S^{11}$ is too small
and noisy to be measured. We return to this in Section \ref{results}.

The diagonal QRC in the T scheme is shown in Figure \ref{fg.scnord}.
It is clear that there is significant $\mu$-dependence on coarser
lattices; in fact for $N_t=4$ we found that $T^2 C_S^{20}(T)/C_S(T)
\simeq -0.096$ at $T/T_c=1.5$. However, this Taylor coefficient vanishes
at the 99\% confidence limit on extrapolation to the continuum. The
conversion from T scheme to Z scheme involves a finite multiplicative
factor $C_S(m_R,T)/C_S(m_R,0)$ which also has an interpretation as
the temperature dependence of the condensate. At $T=1.5T_c$ we find on
extrapolation to the continuum that this factor is $0.26\pm0.01$ for
$m_R$ such that $m_\pi/m_\rho\approx0.6$. With decreasing quark mass
this factor is expected to decrease.

A vanishing continuum limit of the QRC is also found for $m/T_c=0.1$ at
the two other temperatures that we have studied.  The identities in eq.\
(\ref{maxwell}) then imply that $\chi_{uu}$ is insensitive to changes in
the quark mass and $\chi_\pi$ is insensitive to $\mu$ in this range of
temperatures. Since $\chi_{uu}$ agrees with a perturbative evaluation for
$T\ge1.5T_c$, its insensitivity to $m$ can be understood from the fact
that the effective infrared cutoff is given by the Matsubara frequency
$\pi T$ and not by $m$ when $m/T_c<\pi T/T_c$.

\subsubsection{$T<T_c$}

\begin{figure}[!tbh]
   \includegraphics{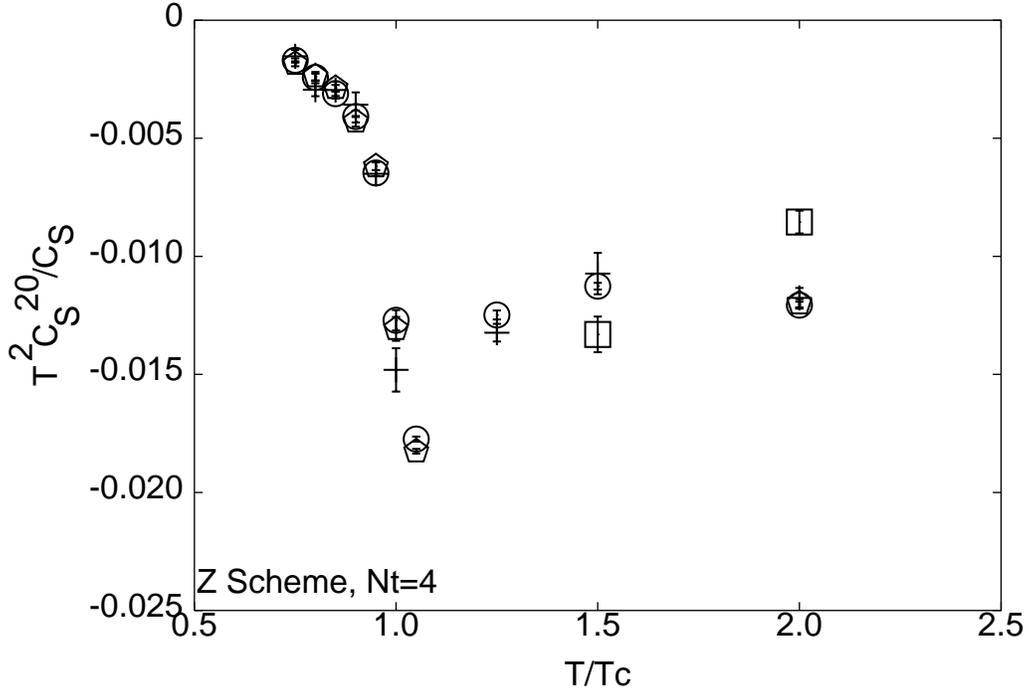}
   \caption{The diagonal QRC as a function of the temperature
            in $N_f=2$ QCD within the Z scheme, for $a=1/4T$ holding
            the renormalized quark mass fixed by fixing the bare mass
            to be $m/T_c=0.1$. Pluses denote results obtained on
            $4\times8^3$ lattices, pentagons on $4\times12^3$ and circles
            on $4\times16^3$. The boxes represent the quenched QCD value.}
\label{fg.nf2}\end{figure}

\begin{figure}[!tbh]
   \includegraphics{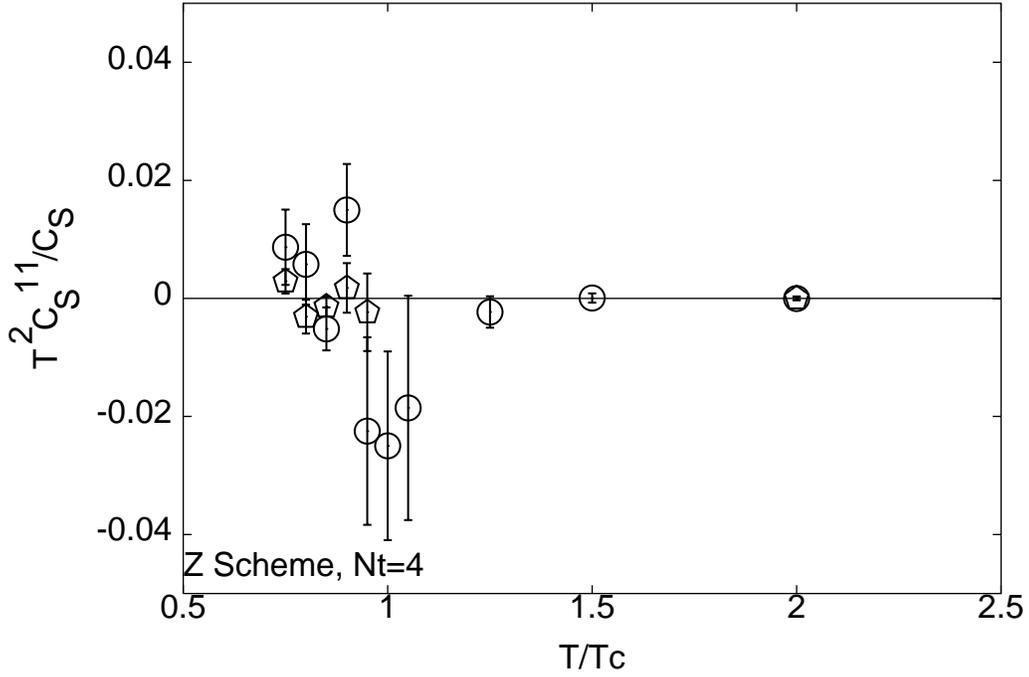}
   \caption{The off-diagonal QRC as a function of the temperature
            in $N_f=2$ QCD within the Z scheme, for $a=1/4T$ holding
            the renormalized quark mass fixed by fixing the bare mass
            to be $m/T_c=0.1$. Pentagons represent results for $4\times12^3$,
            and circles for $4\times16^3$.}
\label{fg.cs11}\end{figure}

\begin{figure}[!tbh]
   \includegraphics{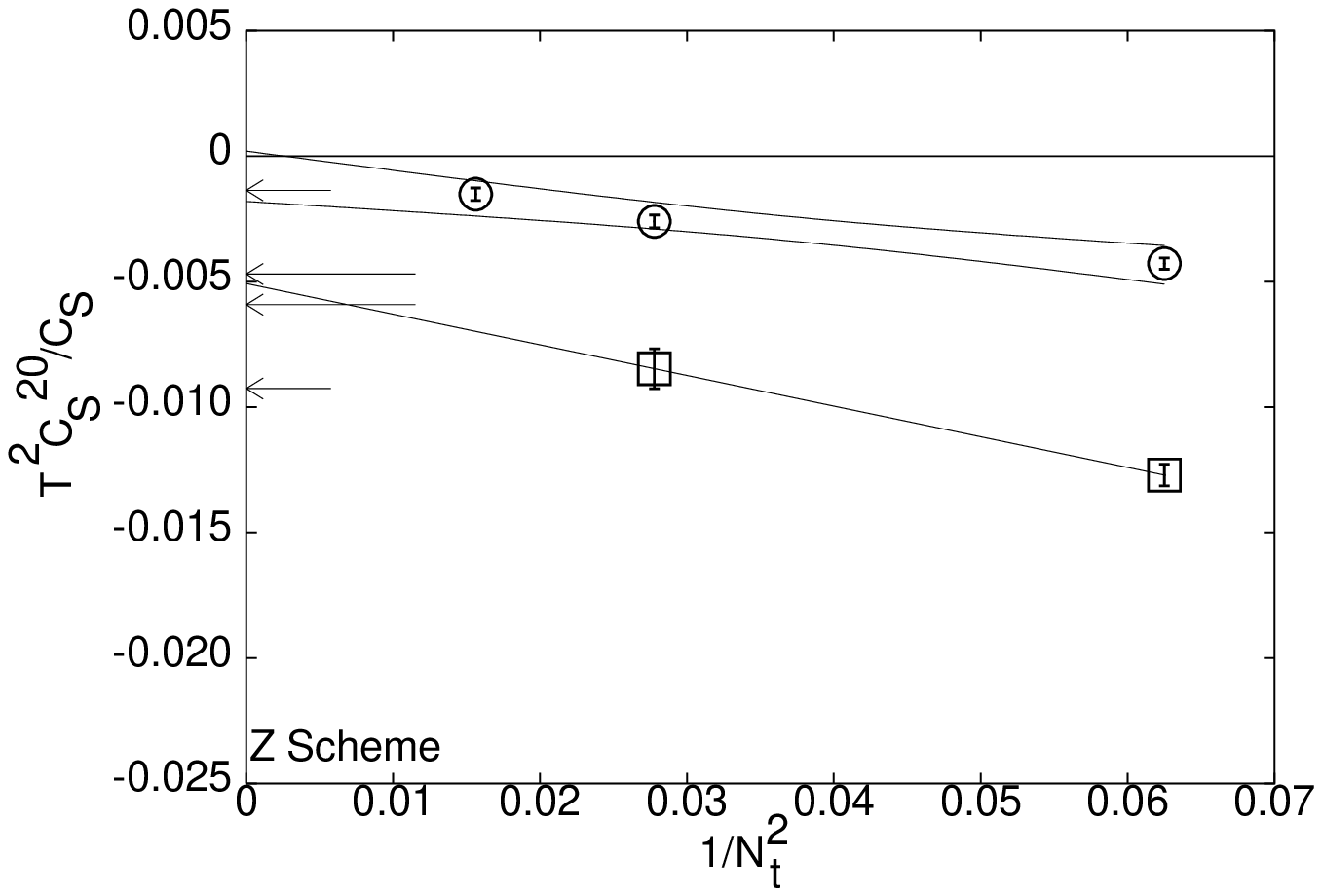}
   \caption{The continuum extrapolation of the diagonal QRC renormalized
      in the Z scheme in $N_f=2$ QCD with bare quark mass of $0.1T_c$.
      The circles denote data for $T=0.9T_c$, and the 3-$\sigma$ band
      for the continuum extrapolation is shown. The boxes represent data
      at $T_c$, and the 1-$\sigma$ (nearer arrows) and 3-$\sigma$
      (further arrows) limits of the extrapolation shown are discussed
      in the text.}
\label{fg.nf2conti}\end{figure}

We also made a series of simulations at fixed $T/T_c=0.75$ for $N_t=6$,
8, 10 and 12 in quenched QCD with the Wilson action. Three hits of
pseudo-heat-bath followed by one Metropolis step made up one composite
sweep. The first 500 sweeps were discarded for thermalization and
subsequently one configuration was saved once every 500 sweeps. Between
35 and 45 such stored configurations were used on each lattice for this
work. Measurements were made with the bare quark mass $m/T_c=0.1$ and
with 50 noise vectors for evaluation of quark traces. In all cases we
used spatial sizes $N_s=2N_t$. From the $T=0$ runs described earlier,
it can be seen that the size of the spatial box corresponds to over 5.5
pion Compton wavelengths (at $T=0$) and hence finite size effects are
under reasonable control.

We found that the continuum limit of both $\langle\cO_2\rangle$ and
$\langle \cO_{11} \rangle$ were consistent with zero within reasonably
small errors. On all but the coarsest lattice, $N_t=6$, the first
expectation value was significantly smaller than the second. As a
consequence $\chi_{ud}$ and $\chi_{uu}$ were equal within errors even at
finite lattice spacing. As shown in Figure \ref{fg.scnorc}, $C_S^{20}$
also vanishes within errors  in the continuum limit, showing that the
vanishing of $\chi_{uu}$ is not an accident of the choice of the quark
mass. The off diagonal QRC was consistent with zero at each $N_t$.

\subsection{Dynamical staggered quarks}

We report the results of a series of computations with two flavours
of dynamical quarks as the temperature varies between $0.75T_c$
and $2T_c$.  The simulations were performed using the R-algorithm
with two flavours of quarks with bare mass of $0.1T_c$, corresponding
to $m_\pi/m_\rho=0.36\pm0.01$ \cite{su3}.  Three lattice sizes were
used in most of the simulations, namely $4\times8^3$, $4\times12^3$ and
$4\times16^3$. The critical coupling $\beta_c$ for $N_t=4$ was determined
in \cite{milc} and the setting of the temperature scale was performed by
methods given in \cite{su3}. The critical coupling for this quark mass
was seen in the peaks of $\chi_\epsilon$, the Wilson line susceptibility,
and even in autocorrelation times \cite{milc,su3}. The critical coupling
is also consistent with those found for neighbouring values of quark
mass in \cite{su3,milc,karsch}.

Lattice computations with dynamical quarks are still too costly for
the continuum limit to be taken easily. However, in order to have
some idea of the continuum limit we have also performed two further
simulations with $6\times12^3$ and $8\times16^3$ lattices at $T=0.9T_c$.
In addition, we have results at $T_c$ on $6\times12^3$ lattices. The
critical couplings at these $N_t$ were determined in \cite{milc2}. It is
impossible to collect a sufficient number of statistically independent
configurations on $N_t=8$ lattices near $T_c$ with current computing
resources due to a rapid increase in autocorrelation times, $\tau$.
In all our simulations autocorrelations of local observables were
monitored and the total run length was at least 50 times the longest,
$\tau_{{\rm max}}$, when measured self consistently over the history. The
measurements used between 50 and 250 configurations, each separated by
$\tau_{{\rm max}}$. We work in the Z scheme since we are interested in
the $T$ dependence of the derivatives.

\begin{figure}[!tbh]
   \begin{center}
   \scalebox{0.5}{\includegraphics{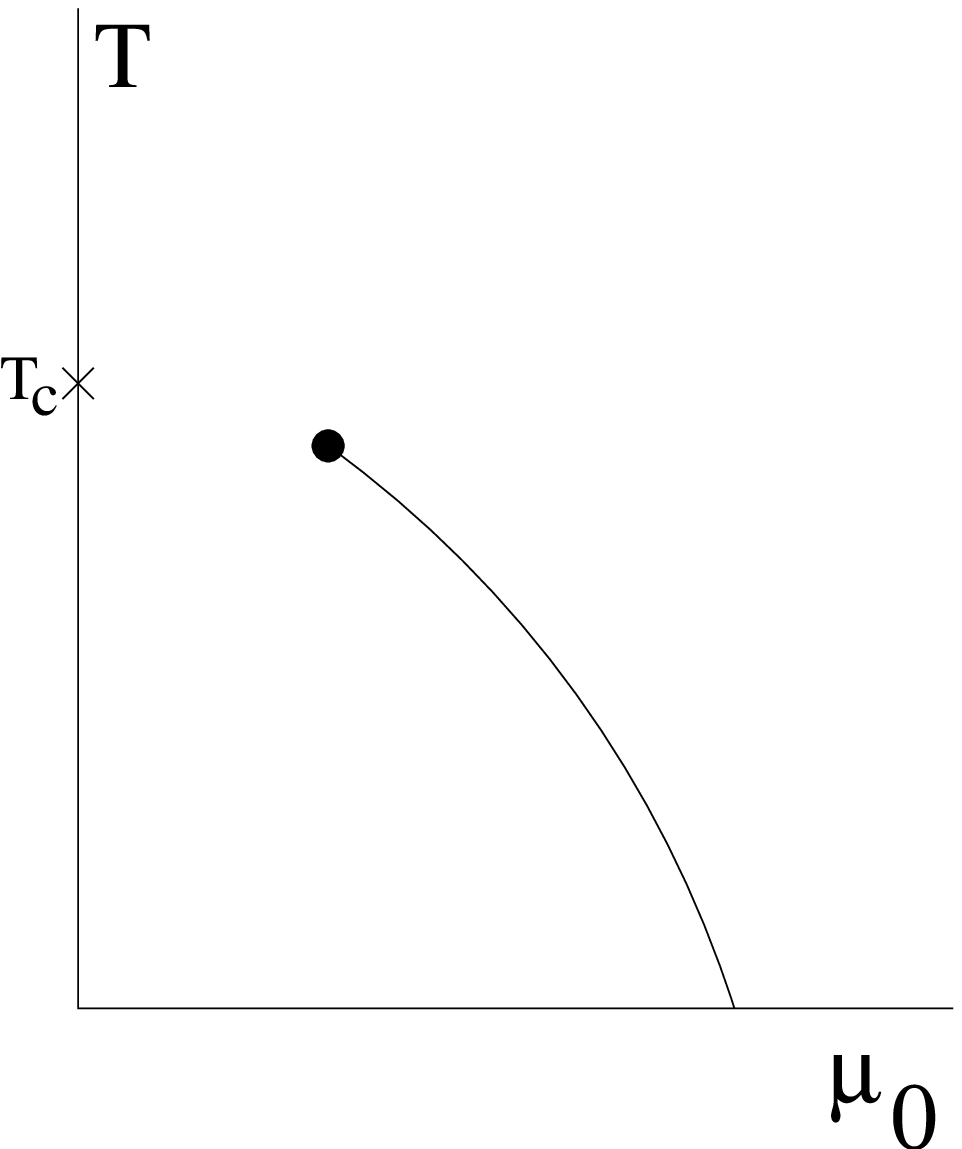}}\hskip2cm
   \scalebox{0.5}{\includegraphics{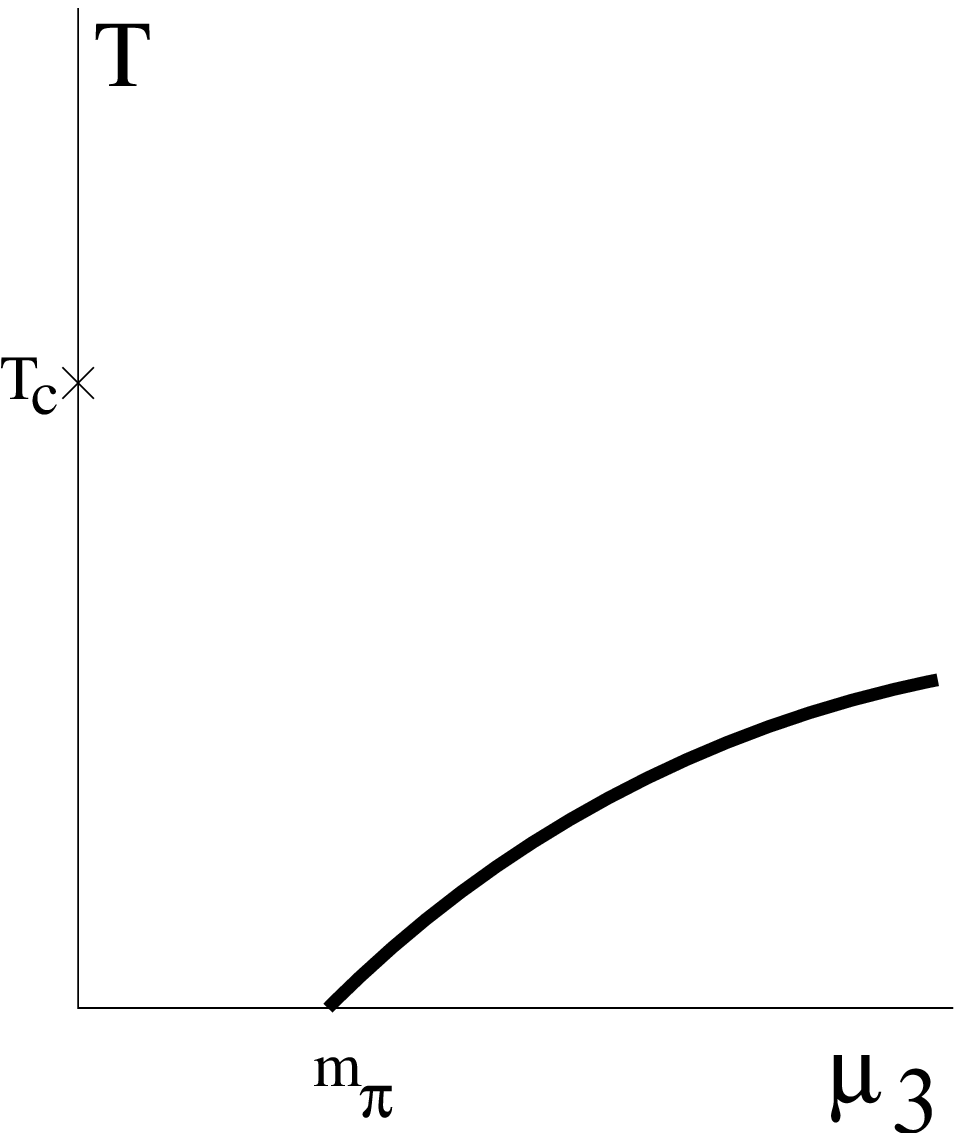}}
   \\
   \scalebox{0.5}{\includegraphics{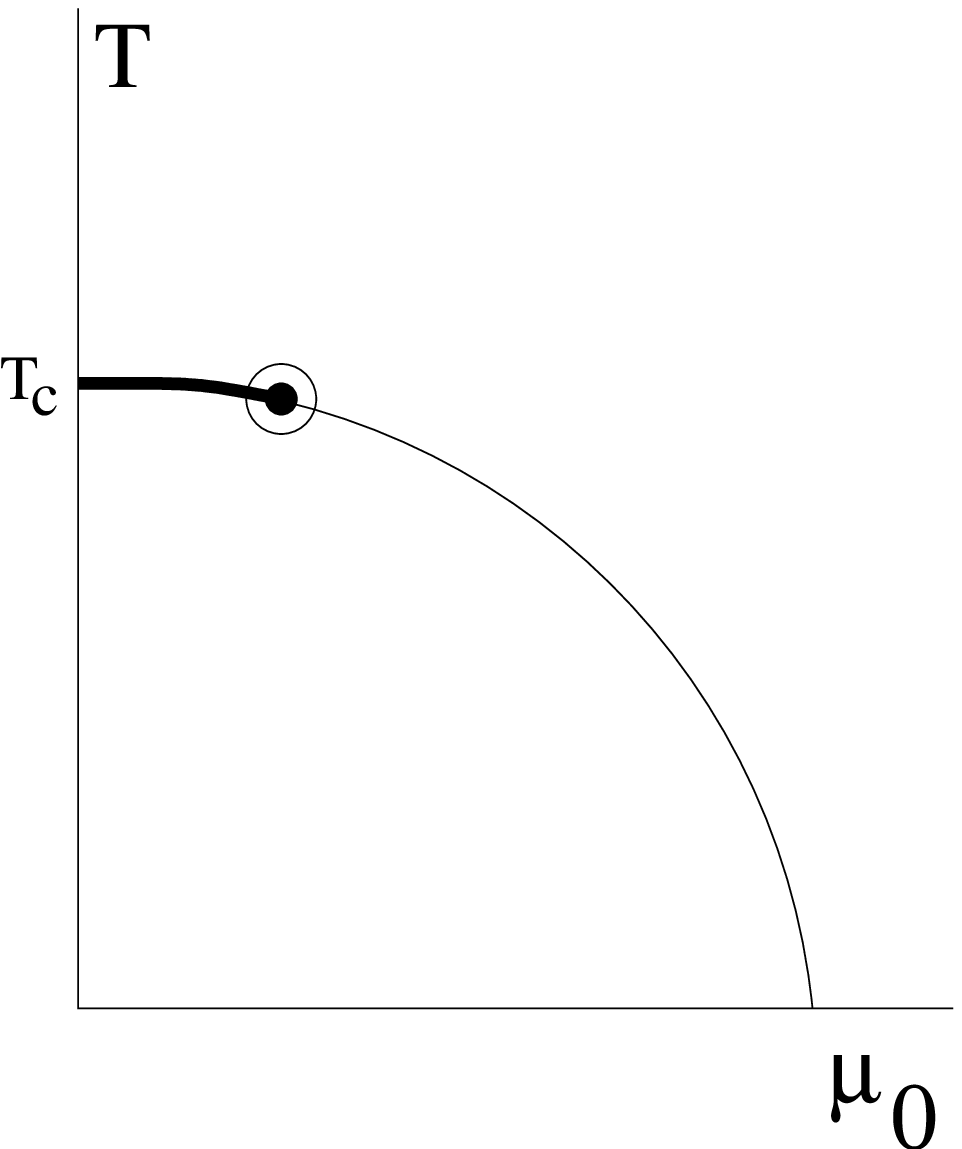}}\hskip2cm
   \scalebox{0.5}{\includegraphics{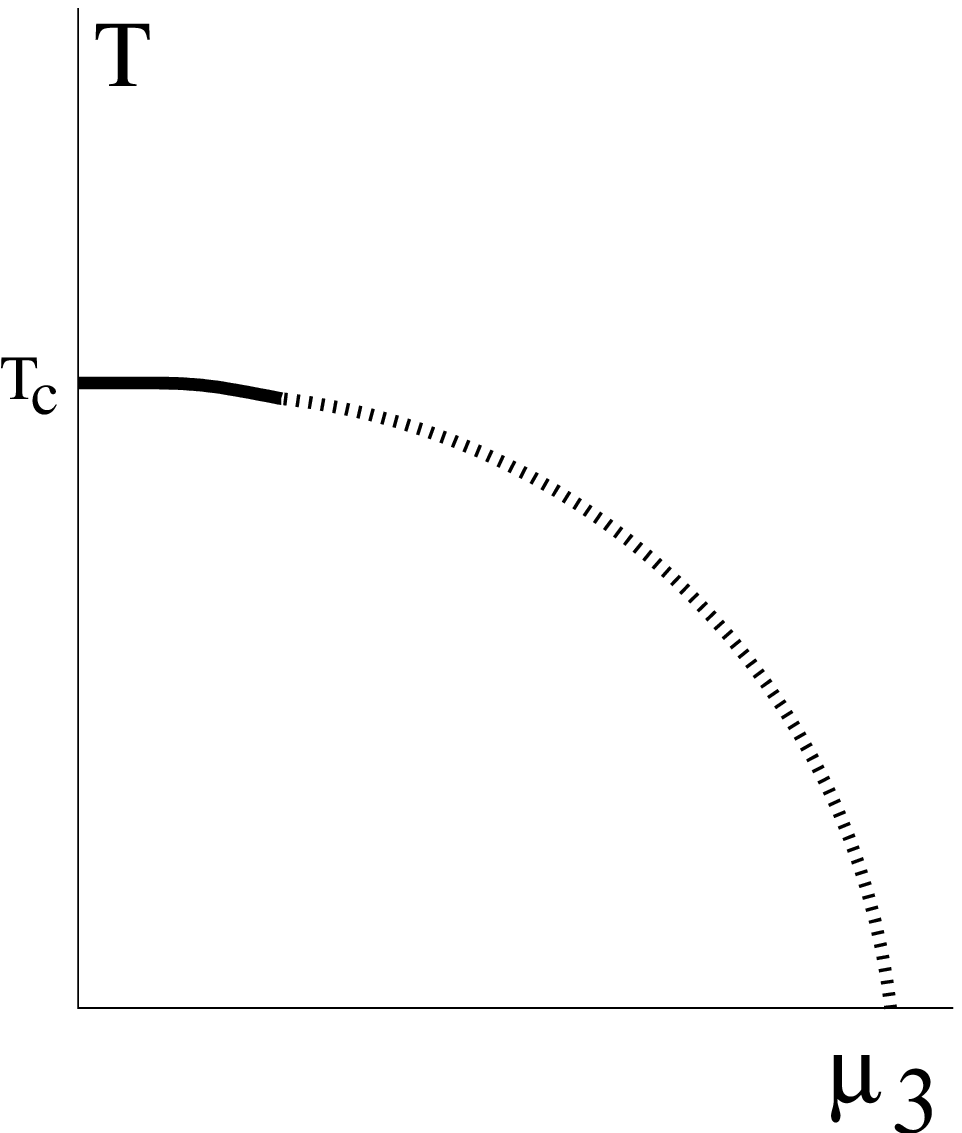}}
   \end{center}
   \caption{The panels on top show the expected phase diagrams in the
      $T$-$\mu_0$ \cite{kr} and $T$-$\mu_3$ \cite{ss} planes for $m_R>0$.
      The panels below show the expectations in the chiral limit, $m_R=0$
      \cite{toublan,nishida}. Thick lines denote second order transitions,
      the tri-critical point is marked with a fancy circle and the dotted
      line shows a part of the critical line which could be a model
      artifact \cite{splnote}. Further interesting phases at large
      $\mu$ are not shown.}
\label{fg.phsdg}\end{figure}

Consistent with our previous observations, $C_S^1$ was seen to vanish
with small errors across the full range of temperatures considered. The
diagonal QRC was negative and differed significantly from zero, as shown
in Figure \ref{fg.nf2}.  For comparison, results from the quenched theory
at the same bare quark mass are also shown. The $T$-dependence is seen to
be a little different; this could reflect either the difference in the
value of $m_\pi/m_\rho$ or be a quenching artifact. The off-diagonal
QRC (see Figure \ref{fg.cs11}) is consistent with zero within small
errors above $T_c$. Near and below $T_c$ larger fluctuations are
seen. Although still consistent with zero at the 3-$\sigma$ level, the
averages increase by several orders of magnitude compared to $T>T_c$,
and seem to be comparable in magnitude to the diagonal QRC.

The results of our preliminary investigation of the continuum limit with
dynamical quarks are shown in Figure \ref{fg.nf2conti}. At $0.9T_c$
the 3-$\sigma$ error band on the extrapolation of the diagonal QRC is
consistent with zero.  At $T_c$ the continuum extrapolation perforce had
to be performed with data from only two lattice spacings. These completely
fix the parameters of a linear extrapolation to the continuum. Error
estimates were obtained by then taking the slope as this given parameter
and estimating the error on the continuum value in the usual way. The
resulting 1-$\sigma$ and 3-$\sigma$ error limits on the extrapolation
are shown in Figure \ref{fg.nf2conti}.  Another estimate is to  take the
extrapolations of the upper end of one error bar and the lower end of
the other. This lies within the 3-$\sigma$ band given.  With all these
caveats, our data currently points to a non-zero continuum extrapolation
of renormalized $C_S^{20}$ at $T_c$.

\section{Discussion}\label{results}

The chiral condensate is a derivative of the free energy density
with respect to quark mass (eq.\ \ref{cond}), and the latter requires
renormalization.  As a result, so does the condensate and its Taylor
coefficients.  We have defined two renormalization schemes for staggered
quarks (eq.\ \ref{scheme}) in order to set up a well defined Taylor
series expansion for the chiral condensate as a function of the chemical
potential $\mu$ (eqs.\ \ref{ctaylor} and \ref{conds}). Diagrammatic
rules for writing down the operators defining the Taylor coefficients
have been given (Figure \ref{fg.chicnd}).  The first order Taylor
coefficient, $C_S^1$ vanishes by symmetry.  The two QRCs are related
through Maxwell relations and chiral Ward identities to all other second
order Taylor coefficients in the chiral sector (eqs.\ \ref{maxwell}).
The diagonal QRC, $C_S^{20}$, has small errors and is easily measurable
at the renormalized quark mass we used.  We found that at finite lattice
cutoff it is negative. Away from $T_c$, both in quenched and $N_f=2$ QCD,
the continuum limit of this quantity vanishes. We have presented evidence
(Figure \ref{fg.nf2conti}) that near $T_c$ in the $N_f=2$ theory it does
not vanish.

This negative peak in $C_S^{20}$ (Figures \ref{fg.nf2} and
\ref{fg.nf2conti}) corresponds to a significant shift of the
condensate, and hence $\chi_\pi$, with chemical potential. The
implications for strangeness production have been pointed out in Section
\ref{sc.maxwell}. Here we discuss how measurements of the diagonal and
off-diagonal QRCs, $C_S^{20}$ and $C_S^{11}$, test current conjectures
about the phase diagram of QCD at small $\mu$ based on the perturbative
expansion or model computations \cite{kr,ss,toublan,nishida}.

The location of $T_c$ is often identified through a peak in the scalar
susceptibility $\chi_\epsilon$ (which is the same as the mass derivative
of the chiral condensate through the Patel Ward identity \cite{patel}). In
the chiral limit this is an unique definition, since the direction of
spontaneous symmetry breaking at low $T$ makes this a massive particle,
which becomes massless precisely at $T_c$. However, at finite quark mass
the symmetry is explicitly broken, $\chi_\pi$ is finite throughout the
low $T$ phase and there is a cross over instead of a critical point along
the $T$ axis at $T_c$ \cite{edwin}, as shown in the top panel of Figure
\ref{fg.phsdg} (see however, \cite{digiacomo}). The cross over temperature
$T_c$ is usually not uniquely defined, in the following sense.  At a phase
transition different physical quantities give exactly the same estimate of
$T_c$. This is no longer true at a cross over.  Thus the slight shift in
the peak of $C_S^{20}$ (Figure \ref{fg.nf2}) compared to $T_c$ previously
defined from peaks in $\chi_\epsilon$ and the Wilson line susceptibilities
\cite{su3,milc} lends some support to the scenario of a cross over.

Furthermore, the observed negative peak in the quadratic response of the
condensate shows that $\chi_\pi$ decreases at finite $\mu$, and hence
pions cannot be the soft mode as one approaches the critical end point.
To the best of our knowledge, this is the first such indication from
lattice QCD.  Unless $C_S^{20}$ changes sign in the continuum, this
argument continues to hold. We have presented evidence that it does,
because the diagonal QRC does not turn positive in the vicinity of $T_c$
(Figure \ref{fg.nf2conti}). Thus support for the standard picture of
the phase diagram for $m_R>0$ is provided by $C_S^{20}$ in two different
ways--- by indicating that at $\mu=0$ there could be a cross over, and by
showing that pion fluctuations decrease as the baryon density increases.

It is interesting to see what else the QRCs can say about the
phase diagram. For $m_R>0$, as $T$ decreases, the Taylor series for
$C_S$ will eventually hit the critical end point in an extrapolation
in $\mu_0$ \cite{kr}, and the critical line in $\mu_3$ \cite{ss,lat} (see
upper panels in Figure \ref{fg.phsdg}). Therefore, at lower $T$ there
must be differences in the two extrapolations. This can only happen if
the off-diagonal QRC becomes significant. We have shown preliminary
evidence for this in Figure \ref{fg.cs11}. This should become easier
to measure with decreasing $m_R$ since $m_\pi$ as well as the chemical
potential at the critical end point are expected to decrease.

In the chiral limit the phase diagram in the $T$-$\mu_0$ plane is
expected \cite{kr} to contain a critical line meeting the $T$ axis at
a critical point $T_c$. It is also expected that this critical line
ends in a tri-critical point where it meets a line of first order phase
transitions emerging from the $\mu_0$ axis (see lower panels in Figure
\ref{fg.phsdg}). The phase diagram in the $T$-$\mu_3$ plane has been
explored in \cite{toublan,nishida}.  A second critical line is expected
to emerge from the $T$ axis, curving down to eventually touch the $\mu_3$
axis \cite{splnote}. Since there is a unique transition along the $T$-axis
for $\mu_0=\mu_3=0$, one expects both critical lines to start from $T_c$.

A Taylor expansion cannot be started at the critical point $T_c$, but
it is possible to make such a series expansion immediately above and
below, and thus track the curvature of the two critical lines. If the
off-diagonal QRC vanishes, then the physics along these two lines is the
same. However, one has massless pions, whereas the other does not. Also,
at lower temperature, one series should give evidence of a tri-critical
point on extrapolation, whereas the other should not. Thus, if the phase
diagrams drawn above are accurate, then they should be reflected by a
non-vanishing value of the off-diagonal QRC. Further, measurements of
$C_S^{20}\pm C_S^{11}$ would give information on the soft modes in these
two directions.

Thus, there are several extensions of this study which would be
interesting to perform in the future.  First, the chiral extrapolation
of $C_S^{20}$ and $C_S^{11}$ obtained at fixed $N_t$ by varying $m_R$
would indicate whether the tri-critical point indeed appears at $m_R=0$.
If the standard picture does hold, then the chiral extrapolation of
$C_S^{20}+C_S^{11}$ and $C_S^{20}-C_S^{11}$ at $T$ just below $T_c$
would give information on the soft modes along the two critical lines.

It is a pleasure to acknowledge discussions with Rajiv Gavai, Edwin
Laermann and Maria-Paola Lombardo.  Part of the computations were carried
out on the Indian Lattice Gauge Theory Initiative's CRAY X1 at the Tata
Institute of Fundamental Research.



\begin{thebibliography}{9}
\bibitem{note}
  In most literature on lattice QCD the sign of the condensate is chosen
  as in eq.\ (\ref{cond}), and $C_S$ is therefore positive. In translating
  our results to the convention with negative $C_S$, care must be taken
  to absorb the sign into the multiplicative renormalization constant,
  so that the renormalizations of mass and chiral condensate differ by
  a sign. Eq.\ (\ref{scheme}) shows that this sign will not appear in
  the expansion of the condensate, but will then appear in the chiral
  Ward identities and the Maxwell relations.
\bibitem{splnote}
   In \cite{nishida} this downward curvature of the critical line in the $T$-$\mu_3$
   plane is due to a lattice artifact called saturation, which occurs when each lattice
   site contains the maximum number of allowed fermions. This is why the downward curve
   has been shown dotted in Figure \ref{fg.phsdg}.
\end{thebibliography}
\end{document}